\documentclass[twocolumn,superscriptaddress,longbibliography,
aps,pra,preprintnumbers]{revtex4-1}

\usepackage{graphicx}
\usepackage{bm}
\usepackage{color}
\usepackage{epstopdf}
\usepackage{amsmath}
\usepackage{amssymb}
\usepackage{epstopdf}

\usepackage[urlcolor=blue,colorlinks=true,citecolor=blue,linkcolor=blue,pdfstartview={FitH},bookmarks=false]{hyperref}

\newcommand{\Imag}{\text{Im}}
\graphicspath{{fig/}{./fig/}{.}}

\begin{document}

\title{Electrostatical formation of the Majorana quasiparticles \\ in the quantum dot--nanoring structure}

\author{Aksel Kobia\l{}ka}
\email[e-mail: ]{akob@kft.umcs.lublin.pl}
\affiliation{Institute of Physics, Maria Curie-Sk\l{}odowska University, \\ 
Plac Marii Sk\l{}odowskiej-Curie 1, PL-20031 Lublin, Poland}

\author{Andrzej Ptok}
\email[e-mail: ]{aptok@mmj.pl}
\affiliation{Institute of Nuclear Physics, Polish Academy of Sciences, \\ ul. W. E. Radzikowskiego 152, PL-31342 Krak\'{o}w, Poland}

\date{\today}

\begin{abstract}
Zero--energy Majorana quasiparticles can be induced at the edge of a low dimensional systems. 
Non--Abelian statistics of this state makes it a good candidate for the realization of quantum computing.
From the practical point of view, it is crucial to obtain an intentional creation and manipulation of this type of bound states.
Here, we show such a possibility in a setup of quantum nanoring in which we specify a quantum dot region via electrostatic means.
States in such quantum dot can lead to the emergence of Andreev and Majorana bound states in investigated system.
We study the differences between those bound states and the possibility of their manipulation.
Moreover, exact calculation method for spectral function has been proposed, which can be used to discuss the bound states influence on band structure of proposed system.
Using this method, it can be shown that the Majorana bound states, induced at the edge of the system, present themselves as a dispersionless zero--energy band.
\end{abstract}


\maketitle

\section{Introduction}
\label{sec.intro}

Kitaev's depiction of the possibility of existence of zero energy Majorana bound states (MBS) in low dimensional setup~\cite{kitaev.01}, ignited the discussion and hopes for creation of quantum computers~\cite{nayak.simon.08,liu.wong.14,sarma.freedman.15}.
MBS properties were extensively studied due to the possibility of application in quantum computing, which is a result of their non--Abelian behaviour~\cite{stanescu.lutchyn.11,alicea.12,leijnse.flensberg.12,stanescu.tewari.13,sarma.freedman.15,lutchyn.bakkers.17}.

Currently, most promising setups where the emergence of MBS is possible are the semiconductor-superconductor nanostructures~\cite{mourik.zuo.12,deng.yu.12,das.ronen.12,finck.vanharlingen.13,deng.vaitiekenas.16,nichele.ofarrell.17,lutchyn.bakkers.17} and ferromagnetic atom chains deposited on the superconductor surface~\cite{nadjperge.drozdov.14,pawlak.kisiel.16,ruby.heinrich.17,feldman.randeria.17,jeon.xie.17}. 
Induced by the proximity effect, the hard superconducting gap~\cite{chang.albrecht.15,gul.zhang.17} leads to the creation of topologically protected zero energy state.

For practical application of MBS in quantum computing, it is necessary to first invent the feasible way for creating and afterwards, a manipulation of MBS.
One of the possible ways is a creation of MBS from two coalescing Andreev bound states (ABS) in a coupled quantum-dot hybrid-nanowire system~\cite{deng.vaitiekenas.16,suominen.kjaergaard.17}.
Another way would require an external, electrostatic control~\cite{ptok.kobialka.17,prada.aguado.17,escribano.levyyeyati.17}, however it requires an additional distinction between trivial and non--trivial zero-energy bound states
~\cite{jeon.xie.17,liu.sau.17,chevallier.szumniak.17,hell.flensberg.17,moore.stanescu.17,cayao.sanjose.17,fleckenstein.dominguez.18}.

\begin{figure}[!b]
\centering
\includegraphics[width=0.75\linewidth]{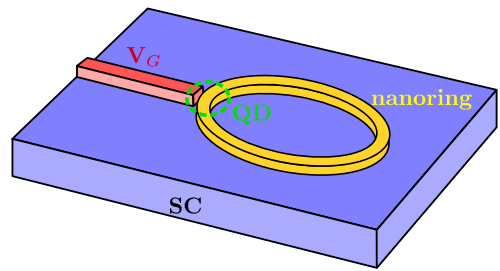}
\caption{
Schematic representation of the described system.
Semiconducting quantum ring (yellow) located at surface of the conventional superconductor (purple).
The quantum dot (QD) can be formed electrostatically by the change of the occupation in some part of the ring (near the gate) by the gate voltage ($V_{G}$).
\label{fig.schem}
}
\end{figure}

For the reasons described above, in this paper, we discuss the possibility of inducing the Majorana quasiparticles in the quantum ring using electrostatic means.
The main idea we propose is shown in  Fig.~\ref{fig.schem} -- using a gate voltage at a part of the semiconducting nanoring located on the superconducting substrate, a quantum dot (QD) region can be created. 
Electrostatic influence of  gate voltage ($V_{G}$) can lead to the emergence MBS in vicinity of the QD.
Similar nanostructures have been experimentally crated already~\cite{fornieri.amado.13}.
We consider a realistic microscopic model and investigate the process of emergence of ABS/MBS and their influence on the band structure of studied system.
Using the Bogoliubov--de~Gennes formalism, we discuss the possibility of \textit{intentional} implementation of zero-energy Majorana bound states. 
Benefiting from the properties of topologically non--trivial setup, we show that such states can be either ABS or MBS, depending on the value of applied voltage $V_{G}$.

This paper is organized as follows.
First, we introduce the theoretical model of our system (Sec.~\ref{sec.model}) and discuss it in the homogeneous case (Sec.~\ref{sec.homo}).
Next, we describe physical properties of the system in the presence of the QD in the real (Sec.~\ref{sec.real}) and momentum space (Sec.~\ref{sec.momentum}).
Finally, we summarize the results in Sec.~\ref{sec.sum}.

\section{Theoretical model}
\label{sec.model}

As mentioned before, we investigate a system in the form of the semiconducting nanoring located on the superconducting substrate (Fig.~\ref{fig.schem}).
This system will be modeled by the one--dimensional Rashba chain with periodic boundary conditions.
Thus, system can be denoted by the Hamiltonian $\mathcal{H} = \mathcal{H}_{0} + \mathcal{H}_{SO} + \mathcal{H}_{QD} + \mathcal{H}_{prox}$. 
The first term describes the mobility of the free electrons in the nanoring:
\begin{eqnarray}
H_{0} = \sum_{ij\sigma} \left( - t \delta_{\langle i,j \rangle} - \left( \mu + \sigma h \right) \delta_{ij} \right) c_{i\sigma}^{\dagger} c_{j\sigma} ,
\end{eqnarray}
where $c_{i\sigma}^{\dagger}$ ($c_{i\sigma}$) is the creation (annihilation) operator of the electron with spin $\sigma$ in {\it i}-th site.
Here $t$ denotes a hopping integral between the nearest-neighbor sites, $\mu$ is a chemical potential and $h$ denotes the magnetic field in the Zeeman form. 
By thorough choice of the gauge symmetry, we can neglect the diamagnetic (orbital) effects~\cite{kiczek.ptok.17}. 
We assume the semiconducting ring, where the spin orbit coupling (SOC) is given by the Rashba term:
\begin{eqnarray}
H_{SO} = - i \lambda \sum_{i \sigma\sigma'} c_{i\sigma}^{\dagger} ( \sigma_{y} )_{\sigma\sigma'} c_{i+1,\sigma'} + h.c. ,
\end{eqnarray}
where $\sigma_{y}$ is the second Pauli matrix and $\lambda$ is the SOC strength.
In the part of the ring, we can form electrostatically the QD.
Using the gate voltage $V_{G}$ we can change the occupation of sites i.e.:
\begin{eqnarray}
\mathcal{H}_{QD} = \sum_{i \in \text{QD} } V_{G} c_{i\sigma}^{\dagger} c_{i\sigma} ,
\end{eqnarray}
where $V_{G}$ plays the role of the additional potential in a part of ring (sites belonging to the QD region).

As a consequence of formation of the nanoring on the superconducting surface, the superconducting gap $\Delta$ is induced by the proximity effect can be described by the BCS-like term:
\begin{eqnarray}
\mathcal{H}_{prox} =  \Delta \sum_{i} \left( c_{i\downarrow} c_{i\uparrow} + c_{i\uparrow}^{\dagger} c_{i\downarrow}^{\dagger}  \right) .
\end{eqnarray}
We assume that superconducting gap depends on the  magnetic field , in the form $\Delta ( h ) = \Delta_{0} \sqrt{ 1 - ( h / h_{c2} )^{2} }$~\cite{liu.sau.17}, where $h_{c2}$ is the critical magnetic field of the bulk system, while $\Delta_{0}$ is the superconducting gap induced in the quantum ring, in the absence of the magnetic field $h$.

\paragraph*{Bogoliubov--de~Gennes formalism.} 
The Hamiltonian $\mathcal{H}$ can be diagonalized by the transformation~\cite{degennes.89}
\begin{eqnarray}
\label{eq.bvtransform} c_{i\sigma} = \sum_{n} \left( u_{in\sigma} \gamma_{n} - \sigma v_{in\sigma}^{\ast} \gamma_{n}^{\dagger} \right) , 
\end{eqnarray}
where $\gamma_{n}$ and $\gamma_{n}^{\dagger}$ are the new quasiparticle fermionic operators, while $u_{in\sigma}$ and $v_{in\sigma}$ are the Bogoliubov--de~Gennes (BdG) eigenvectors.
This transformation leads to the BdG equations in the form $\mathcal{E}_{n} \Psi_{in} = \sum_{j} \mathbb{H}_{ij} \Psi_{jn}$, where
\begin{eqnarray}
\label{eq.bdg}
\mathbb{H} &=& \left(
\begin{array}{cccc}
H_{ij\uparrow} & D_{ij} & S_{ij}^{\uparrow\downarrow} & 0 \\ 
D_{ij}^{\ast} & -H_{ij\downarrow}^{\ast} & 0 & S_{ij}^{\downarrow\uparrow} \\ 
S_{ij}^{\downarrow\uparrow} & 0 & H_{ij\downarrow} & D_{ij} \\ 
0 & S_{ij}^{\uparrow\downarrow} & D_{ij}^{\ast} & -H_{ij\uparrow}^{\ast}
\end{array} 
\right)
\end{eqnarray}
is the Hamiltonian in the matrix form, while 
$\Psi_{in} = \left( u_{in\uparrow} , v_{in\downarrow} , u_{in\downarrow} , v_{in\uparrow} \right)^{T}$. 
Here the matrix elements are: $H_{ij\sigma} = - t \delta_{\langle i,j \rangle} + [ V_{G} \delta_{i \in \text{QD}} - ( \mu + \sigma h ) ] \delta_{ij} $ is the single-particle term, 
$D_{ij} = \Delta \delta_{ij}$ is the on-site superconducting gap, and $S_{ij}^{\sigma\sigma'} = - i \lambda ( \sigma_{y} )_{\sigma\sigma'} \left( \delta_{i+1,j} - \delta_{i-1,j} \right)$ is the SOC term.

We assume the existence of $\mathcal{N}$ sites in the nanoring and $\mathcal{N}_{D}$ sites $( \subset \mathcal{N})$ in the QD region.
In calculations we take $\mathcal{N} = 200$, $\Delta_{0} / t = 0.2$ and $k_{B}T/t = 10^{-5}$.
Other  values will be set at the appropriate times. 
Additionally, for the numerical purposes in some type of calculations (e.g. local density of states or spectral function) we have replaced the Dirac delta function by a Lorentzian $\delta ( \omega ) = \zeta / [ \pi ( \omega^{2} + \zeta^{2} ) ]$ with a broadening of $\zeta / t = 0.001$.

\section{Homogeneous ring}
\label{sec.homo}

In the case of the $\mathcal{N}_{D} = 0$ or $V_{G}/t = 0$, described system is equivalent to the homogeneous ring.
In this case, the Hamiltonian $\mathcal{H}$ can be expressed in the momentum space using the Fourier transform $c_{i\sigma} = \frac{1}{\sqrt{N}} \sum_{\bm k} \exp \left( i {\bm R}_{i} \cdot {\bm k} \right) c_{{\bm k}\sigma}$ as:
\begin{eqnarray}
\nonumber \mathcal{H} &=& \sum_{\bm k} \left( E_{\bm k} - \left( \mu + \sigma h \right) \right) c_{{\bm k}\sigma}^{\dagger} c_{{\bm k}\sigma} \\
&-& i \sum_{\bm k} \mathcal{L}_{\bm k} \sum_{\sigma\sigma'} c_{{\bm k}\sigma}^{\dagger} \left( \sigma_{y} \right)_{\sigma\sigma'} c_{{\bm k}\sigma'} \\
\nonumber  &+& \Delta \sum_{\bm k} \left( c_{{\bm k}\uparrow} c_{-{\bm k}\downarrow} + h.c. \right)
\end{eqnarray}
where $c_{{\bm k}\sigma}^{\dagger}$ ($c_{{\bm k}\sigma}$) is the creation (annihilation) operator of electron with momentum ${\bm k}$ and spin $\sigma$.
Here $E_{\bm k} = - 2 t \cos ( k )$ is the dispersion relation for non--interacting electron with momentum ${\bm k}$, while $\mathcal{L}_{\bm k} = 2 \lambda \sin ( k )$ denotes the spin-orbit coupling.

\begin{figure}[!t]
\centering
\includegraphics[width=0.8\linewidth]{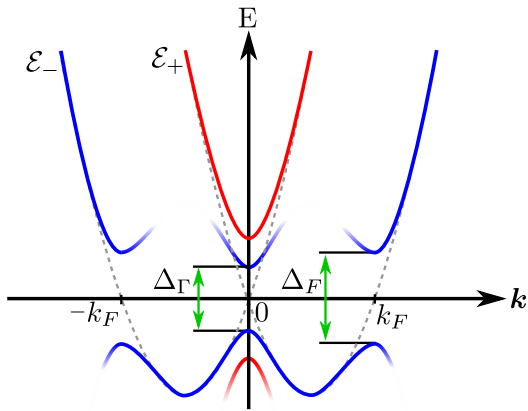}
\caption{
Schematic representation of the band structure of homogeneous system for the superconducting state in the presence of the external magnetic filed.
Blue and red solid line represent lower and upper Rashba bands, respectively. 
Gray dashed parabolas represent the band structure without the influence of magnetic field.
\label{fig.schem_band}
}
\end{figure}

\paragraph*{Band structure.}
Let us shortly describe main properties of the band structure in a homogeneous case (Fig.~\ref{fig.schem_band}).
As a consequence of the existence of SOC ($\lambda > 0$) in the normal state ($\Delta_{0} = 0$) band structure is represented by two shifted parabolas (gray dashed line), crossing the Fermi level ($E = 0$) at the ${\bm k} = 0$ and ${\bm k} = k_{F}$.
Turning on the superconductivity opens the gap around these two points (marked by $\Delta_{\Gamma}$ and $\Delta_{F}$, respectively).
In the absence of the magnetic field ($h = 0$, not shown), both gaps are equal ($\Delta_{\Gamma} = \Delta_{F} = \Delta_{0}$).
Increase of the magnetic field leads to shift of the spin up/down energy levels and decreases the gap at ${\bm k} = 0$ [then $\Delta_{\Gamma} = \Delta ( h )  - h$].
If the superconducting gap is large enough increasing the magnetic field leads to the closing the gap at ${\bm k} = 0$ at some critical magnetic field $h_{c}$ and reopens the new {\it topological gap}.
Closing trivial (superconducting) gap and reopening the new topological gap as the magnetic field $h$ surpasses the $h_{c}$ threshold, is associated with the topological phase transition.
Band inversion is a consequence of transition from trivial to non--trival topological phase~\cite{hasan.kane.10,bansil.lin.16} (which will be shown  in Sec.~\ref{sec.momentum}).

\paragraph*{Topological phase transition.}
For the homogeneous system we can calculate the condition for the topological phase transition (from the trivial to non-trivial phase).
Topological phase transition occurs, when the gap of the system closes, which is equivalent to the condition~\cite{sato.fujimoto.09,sato.takahashi.09,sato.takahashi.10}:
\begin{eqnarray}
&& \left( E_{\bm k} - \mu \right)^{2} + h^{2} + | \mathcal{L}_{\bm k} |^{2}  + | \Delta |^{2} \\
\nonumber &=& 2 \sqrt{ \left( E_{\bm k} - \mu \right)^{2} | \mathcal{L}_{\bm k} |^{2} + \left( \left( E_{\bm k} - \mu \right)^{2} + | \Delta |^{2} \right) h^{2} } 
\end{eqnarray}
which give~\cite{sato.06}:
\begin{eqnarray}
\left( E_{\bm k} - \mu \right)^{2} + | \Delta |^{2} = h^{2} + | \mathcal{L}_{\bm k} |^{2} , \quad \Delta \mathcal{L}_{\bm k} = 0 .
\end{eqnarray}
Second condition is met by ${\bm k} = 0$ and ${\bm k} = \pm \pi$, which are two time--reversal--invariant momenta for one--dimensional system~\cite{moore.balents.07}.
By inserting those values to the first equation, we get:
\begin{eqnarray}
\label{eq.qpt_hc} h_{c} = \sqrt{ \left( 2 t \pm \mu \right)^{2} + | \Delta |^{2} } ,
\end{eqnarray}
where  $\Delta = \Delta ( h )$.
This value of the magnetic field describes the critical value in which the trivial energy gap closes and new topological gap is reopened.

\section{Real space description}
\label{sec.real}

To study our system we will use the spin dependent local density of states (LDOS) $\rho_{i\sigma} ( \omega ) =  -\frac{1}{\pi} \Imag \langle \langle c_{i\sigma} | c_{i\sigma}^{\dagger} \rangle\rangle$, which by using transformation~(\ref{eq.bvtransform}) can be expressed in well known form~\cite{matsui.sato.03}:
\begin{eqnarray}
\nonumber \rho_{i\sigma} ( \omega ) = \sum_{n} \left[ | u_{in\sigma} |^{2} \delta \left( \omega - \mathcal{E}_{n} \right) + | v_{in\sigma} |^{2} \delta \left( \omega + \mathcal{E}_{n} \right) \right] , \\ \label{eq.ldos}
\end{eqnarray}
while the total density of states (DOS) is given by $\rho ( \omega ) = \sum_{i\sigma} \rho_{i\sigma} ( \omega )$.
Those physical quantities are measurable by the scanning tunneling microscope (STM)~\cite{hofer.foster.03,wiesendanger.09,oka.brovko.14} and are equivalent to the differential conductance $G_{i} = dI_{i} (V) / dV$ measurement~\cite{figgins.morr.10,chevallier.klinovaja.16,stenger.stanescu.17}.

\begin{figure}[!b]
\centering
\includegraphics[width=\linewidth]{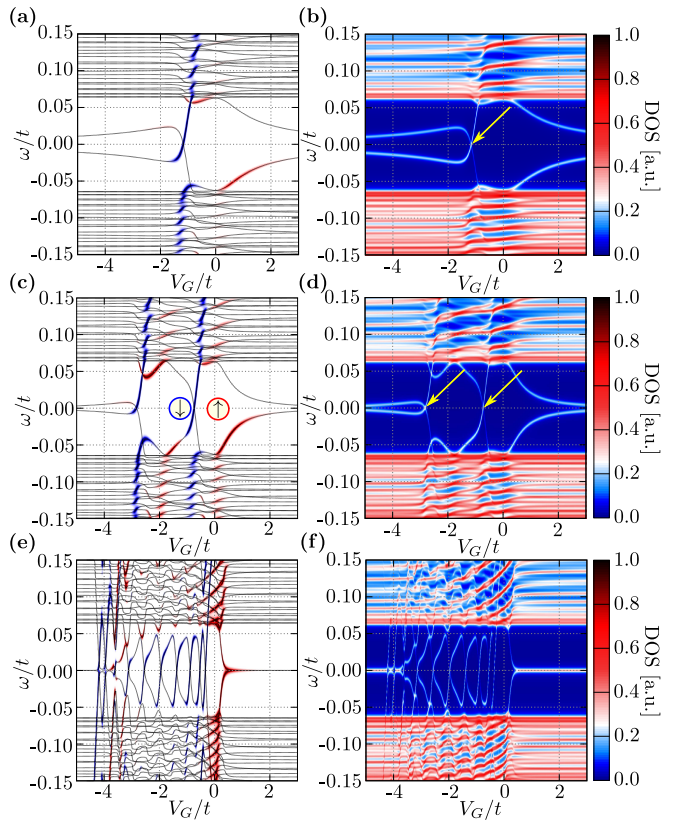}
\caption{
Low energetic spectrum (left row) and density of states (right row) of the system for different number of sites $\mathcal{N}_{D}$ in the quantum dot region (from top to bottom the $\mathcal{N}_{D}$ is equal to 1, 2 and 10).
Background color in the left row denotes contribution to the quantum dot states with spin up (red) and down (blue) character to the system.
Results in the presence of the magnetic field ($h = 0.3 t > h_{c}$).
\label{fig.nd_dos}
}
\end{figure}

\begin{figure}[!b]
\centering
\includegraphics[width=\linewidth]{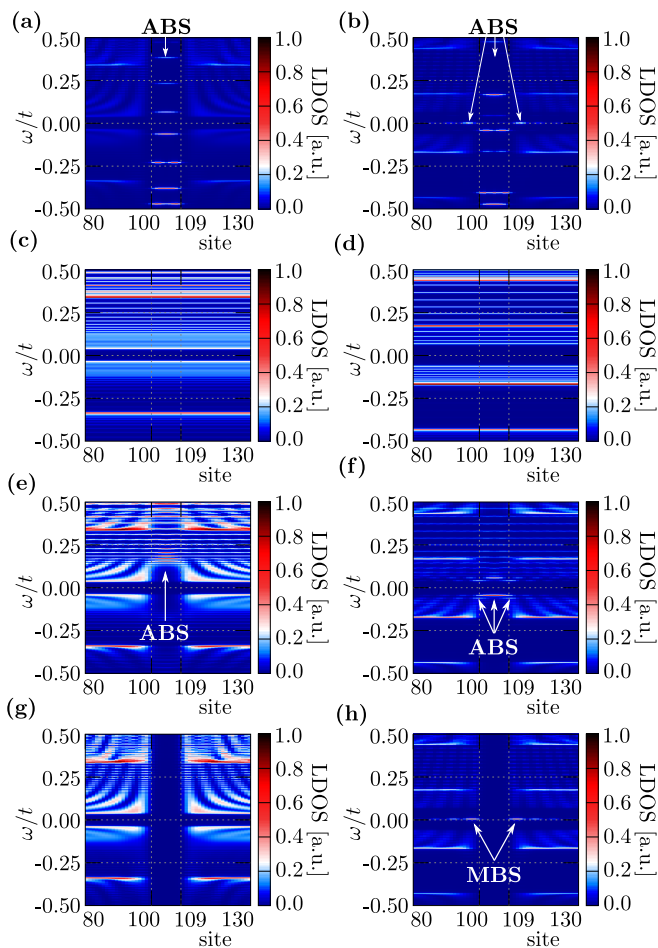}
\caption{
Local density of states (LDOS) around the quantum ring region (for different value of $V_{G}/t$ equal $-4.0$, $0.0$, $0.2$ and $1.0$, from top to bottom).
Quantum dot is located within $100$ to $109$ sites ($\mathcal{N}_{Q} = 10$ ).
Results for non-topological (trivial) phase ($h = 0.15 t < h_{c}$) and topologically non-trivial phase ($h = 0.3 t > h_{c}$) in left and right row, respectively.
\label{fig.ldos_ring}
}
\end{figure}

Increase of the gate voltage $V_{G}$ in the QD region ($\mathcal{N}_{D} > 0$), leads to the emergence of bound states on the newly created defect in the periodic structure.
Its form and influence on the system is dependent on $V_{G}$ and $\mathcal{N}_{D}$.
Examples of numerical results are shown in Fig.~\ref{fig.nd_dos}.
Initially, degenerated ABS split under the influence of Zeeman effect.
The in--gap bound states are visible in the form of the ABS in the DOS structure (right panels), which number depends on the $\mathcal{N}_{D}$.
Energy of the ABS is a non-trivial function of gate potential $V_{G}$.
For some value of $V_{G}$ (yellow arrows) two ABS can coalesce at the zero--energy.
This value of $V_{G}$ strongly depends on the available QD energy levels (cf. panels in one row). 
As we know, the MBS have the spin--polarization similar to the system majority spin (i.e. $\uparrow$)~\cite{sticlet.bena.12,maska.domanski.17}.
From this,  zero--energy bound states crossing are not MBS, because of the apparent opposite polarization (blue colors at the background of the left panels).
The QD states with the dominant spin--up character are well hybridized with the rest of the ring spin--up state by the spin--conserving hopping ($t$)~\cite{ptok.kobialka.17,guigou.sedlmayr.16,kobialka.ptok.18}.
As a consequence of this, spin--up QD states does not cross  the Fermi level (red colors at the background of the left panels).
Situation looks different for a case of the minority spin QD states (blue colors in left panels).
This spin-down QD states hybridization with the rest of ring spin-up states is possible only by the spin--flip hopping $\lambda \ll t$.
As a consequence, we observe the crossing at the Fermi level.
Properties described above are well visible in the numerical results, for the case of QD with $\mathcal{N}_{D} = 10$ sites [Fig.~\ref{fig.nd_dos}(e) and (f)].
In the region where the QD energy levels cross the gap, we observe several crossings of the Fermi level by the spin--down QD dot levels ($V_{G} \sim - 4 t$) and absence of the spin-up ABS ($V_{G} \sim 0 t$).
In relation to the system without periodic boundary conditions e.g. QD--hybrid nanowire system~\cite{ptok.kobialka.17}, we do not observe MBS independent of the QD levels MBS in the system for $- 4 < V_{G} / t < 0$.

A relatively simple way to show the difference between ordinary ABS and non-trivial MBS is the investigation of the LDOS spectrum around the QD region.
Numerical results are shown in Fig.~\ref{fig.ldos_ring}, for a case of the trivial and non-trivial phase (at left and right panels, respectively).
When the spin-down QD energy levels cross the gap region [e.g. $V_{G}/t = -4$ shown in Fig.~\ref{fig.ldos_ring}(a) and (b)], we can observe the ABS inside the QD region.
Additionally, in the non-trivial topological phase, we can observe the ABS localized outside of the QD region [Fig.~\ref{fig.ldos_ring}(b)]. 
This bound states are located around the Fermi level [cf. with Fig.~\ref{fig.nd_dos}(e) where eigenvalues \textit{avoided} cross the zero--energy level].
In the case of the homogeneous system ($V_{G}/t = 0$), we see uniform lines independent of the external magnetic field $h$[Fig.~\ref{fig.ldos_ring}(c) and (d)].
The positive gate potential $V_{G}$ leads to the interplay between the spin-up QD and ring levels. 
As a consequence of the hybridization described in previous paragraph, we do not observe low energetic ABS localized outside the QD region [Fig.~\ref{fig.ldos_ring} panels (c) and (d)].
The ABS localized inside the QD region are associated with the $V_{G}$--shifted spin-up QD energy levels within superconducting gap.

For sufficiently large value of $V_{G}$, the QD energy levels are located above the gap region [Fig.~\ref{fig.ldos_ring}(g) and (f)].
In trivial phase, zero--energy bound states are not observed in vicinity of the QD region [Fig.~\ref{fig.ldos_ring}(g)].
In contrary, when the system is in the non--trivial topological phase ($h > h_{c}$) the bound states are realized in the form of the MBS localized outside the QD region [Fig.~\ref{fig.ldos_ring}(f)]~\cite{maska.gorczyca.17}.
Those properties can be explained by the interplay of the {\it defect} (in the QD form) with the rest of the system and additionally its influence on the band structure, what will be discussed in the Sec.~\ref{sec.momentum}.

\begin{figure}[!t]
\centering
\includegraphics[width=\linewidth]{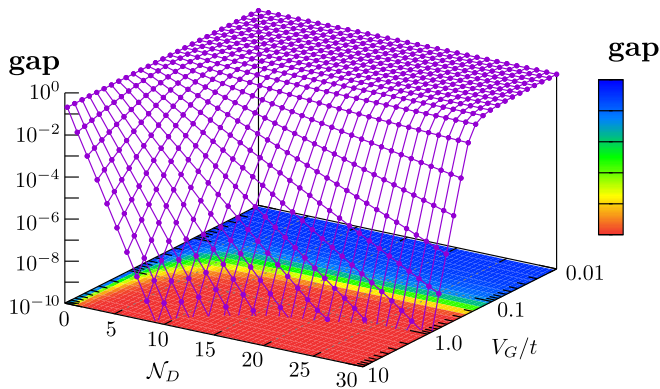}
\caption{
The phase diagram of gate potential $V_{G}$ versus number of sites $\mathcal{N}_{D}$ in the quantum dot region.
Z-axis and background colors shown the gap between two eigenvalues closest to the Fermi level.
Results for $h/t = 0.3$.
\label{fig.gap_nd_vg}
}
\end{figure}

Increasing the gate potential $V_{G}$ to relatively large value with respect to the rest of the ring states (in our case below $-4 t-h$ and above $h$), leads to occurrence of the new boundary at the QD--ring connection.
Absence of the ring  energetic levels in the QD regions leads to the emergence the (initially Andreev and finally Majorana) bound states~\cite{dassarma.sau.12,albrecht.higginbotham.16} localized in the ring region.
Indeed, it is {\it a~priori} seen in the dependence of the gap between two bound states closest to zero-energy, as a function of gate potential $V_{G}$ and number of sites in the QD region $\mathcal{N}_{D}$ (Fig.~\ref{fig.gap_nd_vg}).
Here, $\mathcal{N}_{D}$ correspond to the distance between the bound states localized outside of the QD region.
The bound states can be realized easier in the larger QD, as the bound state wavefunction overlap would be lesser and therefore annihilation of MBS would take longer.
Moreover, we can observe ABS--MBS crossover around $V_{G}/t \sim 0.5$. 
An increase of the $\mathcal{N}_{D}$ leads to decrease of the $V_{G}$ for which the ABS form the MBS.
This scheme is essential for possible application of studied system in creation of MBS and its manipulation using electrostatic and magnetic means, and it can prove beneficial for the quantum computing.

\begin{figure}[!b]
\centering
\includegraphics[width=\linewidth]{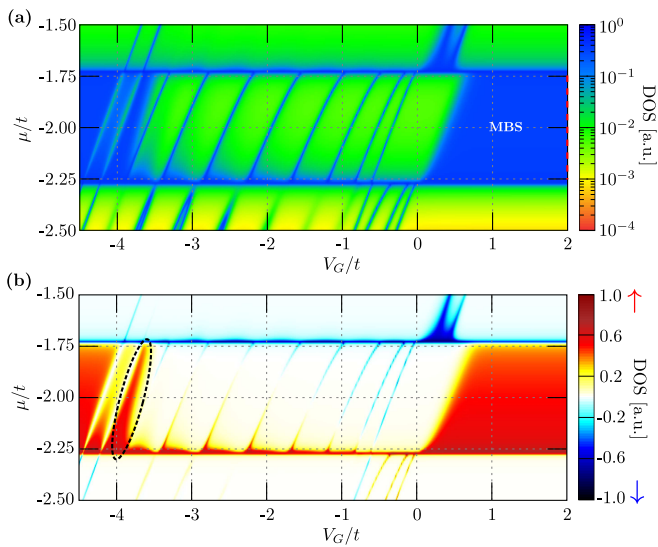}
\caption{
Zero energy total DOS (a) and polarization of the zero-energy state (b) for different values of the gate potential $V_{G}$ and chemical potential $\mu$.
Results for $\mathcal{N}_{D} = 10$. 
(c-f) Crossection of the DOS (for chosen value of $V_{G}$ shown in panel (a) by red arrows) as a function of the chemical potential $\mu$. 
The gate potential $V_{G}/t$ is $-4.3$ (c), $-3.75$ (d), $0.0$ (e), $2.0$ (f).
Result in the presents of the magnetic field ($h = 0.3 t > h_{c}$).
\label{fig.dos_mu}
}
\end{figure}

Additional information about realized type of the bound state in the system can be given by the total DOS and polarization at the zero energy level (Fig~\ref{fig.dos_mu}).
Analysis of the zero--energy DOS [Fig~\ref{fig.dos_mu}(a)] show places (in the form of inclined blue lines) where the zero--energy bound  states are created as the two ABS levels crossing the Fermi level.
However, comparing those results with the polarization at the zero--energy level [difference between zero energy DOS for spin up and down, presented at Fig~\ref{fig.dos_mu}(b)] shows that the real MBS emerge only for some value of $V_{G}$ (dark red regions).
Therefore a spin resolved measurements can be essential in distinguishing the MBS from normal, zero--energy bound states.
Indeed, as we have stated previously, only when the QD dot filling $n_{D} = 1/\mathcal{N}_{D} \sum_{i \in \text{QD}, \sigma} \langle c_{i\sigma}^{\dagger} c_{i\sigma} \rangle$ is significantly different from the rest of the ring (i.e. for gate potential $V_{G}$ smaller than $-4t-h$ for the fully filled $n_{D} \simeq 2$ or bigger than $ h$ for the empty QD $n_{D} \simeq 0$), the MBS can occur in our system.
Behavior in this regime is a consequence of the strong hybridization between the similar--spin QD and ring states.
Moreover, for small  QD filling, the MBS can emerge from the coalescence of the ABS localized outside the QD region [e.g. for parameters shown by the black dashed ellipse in the Fig~\ref{fig.dos_mu}(b)].
It is a consequence of weak mixing betweem the spin--down QD energy levels and the spin--up ring energy levels.

\begin{figure}[!b]
\centering
\includegraphics[width=\linewidth]{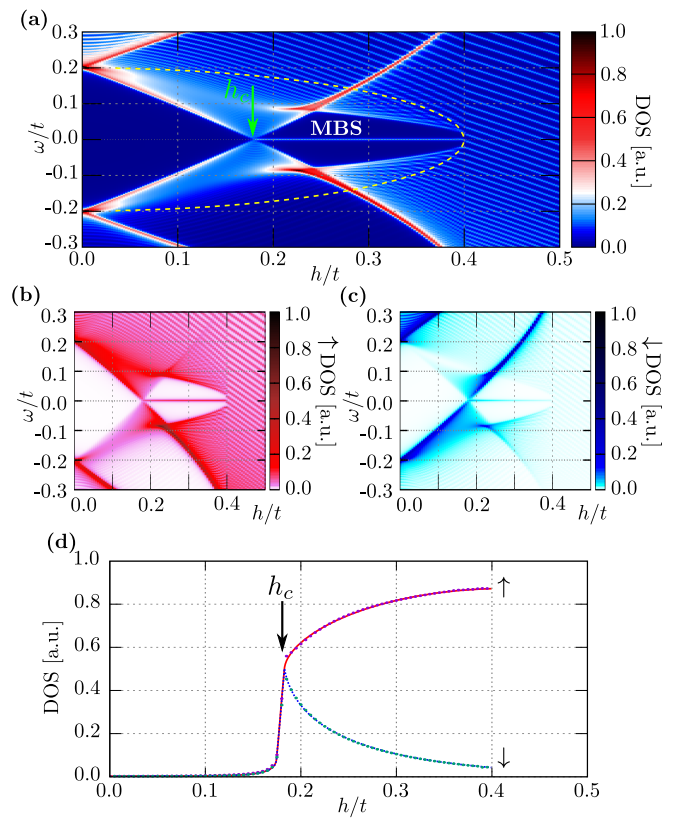}
\caption{
DOS (a) and spin-dependent DOS (b-c) as a function of the external magnetic field $h$.
(d) Contribution of the spin-dependent DOS to the MBS.
Yellow line shows value of magnetic field dependent superconducting gap $\Delta(h) = \Delta_{0} \sqrt{1 - ( h / h_{c2} )^{2} }$, while green arrow defines topological phase transition from trivial to non-trivial phase, for magnetic field  $h_{c}$.
Results for $\mathcal{N}_{D} = 10$ and $V_{G}/t = 2$.
\label{fig.dos_em}
}
\end{figure}

In relation to  properties described above , it is important to formulate a description of how the MBS are composed of the spin up and down particles.
Analysis of the total DOS as a function of the magnetic field $h$ for fixed $\mathcal{N}_{D}$, show typical character [Fig.~\ref{fig.dos_em}(a)].
Increase of $h$ leads to the closing of the trivial gap (associated with $\Delta_{\Gamma}$ on Fig.~\ref{fig.schem_band}), and reopening the new topological gap at the $h_{c}$ [given by Eq.~(\ref{eq.qpt_hc}), as marked by yellow arrow].
For $h_{c} < h < h_{c2}$, we can find signs of the zero-energy MBS.
Analysis of the spin--dependent DOS [Fig.~\ref{fig.dos_em}(b) and (c)] clearly shows the band inversion at the $h_{c}$.
In the topological phase, spectral weight corresponding to the spin--down state is mostly located above the Fermi level [Fig.~\ref{fig.dos_em}(c)]. 
Moreover, contribution of the spin--up and --down states in the MBS changes [Fig.~\ref{fig.dos_em}(d)]. 
As we can see, the MBS state is realized above $h_{c}$ as a zero--energy bound state with unequal composition of the spin--up and --down, due to its emergent nature.
Additionally, ratio between spin up and down components of the MBS increases with $h$.
It should be noted that the minority spin component is always non--zero. 
It means that the MBS  have ``spin'' pointed at the magnetic field direction, which is in agreement with previous theoretical study~\cite{sticlet.bena.12,maska.domanski.17} and experimental measurements~\cite{jeon.xie.17}.
It should be mentioned, that spin properties of the MBS should be always included in the effective {\it minimal} model of this type of bound states~\cite{prada.aguado.17,deng.vaitiekenas.17}.

\section{Momentum space description}
\label{sec.momentum}

The LDOS show us how inhomogeneity leads to the localization of the Andreev and/or Majorana bound states in the real space.
In similar way, in the momentum space we can define the spectral function which can show how the same defects influence the band structure of our system~\cite{xu.chiu.17}.
The spectral function, similarly to the LDOS, is a measurable quantity.
In this case the angle-resolved photoemission spectroscopy (ARPES) technique~\cite{damascelli.hussain.03}, can be used for visualization of the band structure, even in the case of the nanostructures~\cite{snijders.weitering.10}.
As mentioned previously, in the system with the periodic boundary conditions, the momentum ${\bm k}$ is a well defined quantity, even in the presence of the defects.

To investigate the influence of QD on the band structure of our system, we will use the spectral function $\mathcal{A}_{{\bm k}\sigma} ( \omega ) = - \frac{1}{\pi} \Imag \mathcal{G}_{{\bm k}{\bm k}\sigma} ( \omega )$~\cite{mayr.alvarez.06}. 
To do this, we must first define the Green functions in the real space $G_{ij\sigma} (\omega) = \langle\langle c_{i\sigma} | c_{j\sigma}^{\dagger} \rangle\rangle$ and the Green functions in the momentum space $\mathcal{G}_{{\bm k}{\bm l}\sigma} (\omega) = \langle\langle c_{{\bm k}\sigma} | c_{{\bm l}\sigma}^{\dagger} \rangle\rangle$.
The mutual relation between $G_{ij\sigma}$ and $\mathcal{G}_{{\bm k}{\bm l}\sigma}$ functions is  given {\it explicite} in the periodic system by the Fourier transformation:
\begin{eqnarray}
\label{eq.ftgf} G_{ij\sigma} = \frac{1}{N} \sum_{{\bm k}{\bm l}} \exp \left( i {\bm k} \cdot {\bm R}_{i} \right) \exp \left( - i {\bm l} \cdot {\bm R}_{j} \right) \mathcal{G}_{{\bm k}{\bm l}\sigma} .
\end{eqnarray}
Additionally, using transformation~(\ref{eq.bvtransform}) we can $G_{ij\sigma}$ rewrite in the form:
\begin{eqnarray}
\nonumber G_{ij\sigma} = \sum_{n} \left( u_{in\sigma} u_{jn\sigma}^{\ast} \langle\langle \gamma_{n} | \gamma_{n}^{\dagger} \rangle\rangle + v_{in\sigma}^{\ast} v_{jn\sigma} \langle\langle \gamma_{n}^{\dagger} | \gamma_{n} \rangle\rangle \right) . \\
\end{eqnarray}
Then, from Eq.~\ref{eq.ftgf}, the spectral function is given as:
\begin{eqnarray}
\label{eq.funspec} & \mathcal{A}_{{\bm k}\sigma} ( \omega ) & = \frac{1}{N} \sum_{ij} \exp \left( - i {\bm k} \cdot \left( {\bm R}_{i} - {\bm R}_{j} \right) \right) \times \\
\nonumber && \sum_{n} \left( u_{in\sigma} u_{jn\sigma}^{\ast} \delta \left( \omega - \mathcal{E}_{n} \right) + v_{in\sigma}^{\ast} v_{jn\sigma} \delta \left( \omega + \mathcal{E}_{n} \right) \right) .
\end{eqnarray}
From solution of the BdG equations~(\ref{eq.bdg}) in real space, we can  numerically find band structure in the form of the spectral function $\mathcal{A}_{{\bm k}\sigma} ( \omega )$.
Method described above can be also generalized to the finite size system (without periodic boundary condition) e.g. in the nanowire form.

\begin{figure}[!b]
\centering
\includegraphics[width=\linewidth]{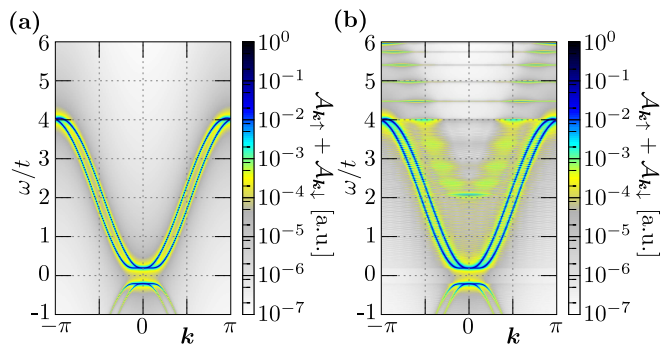}
\caption{
The spectral functions $\mathcal{A}_{{\bm k}\uparrow} + \mathcal{A}_{{\bm k}\downarrow}$ of the system in the absence of the magnetic field for the gate potential $V_{G}$ equal $0 t$ (a) and $2 t$ (b).
Results for a case of the QD with $\mathcal{N}_{D} = 10$ sites.
\label{fig.band_h0}
}
\end{figure}

\begin{figure}[!t]
\centering
\includegraphics[width=\linewidth]{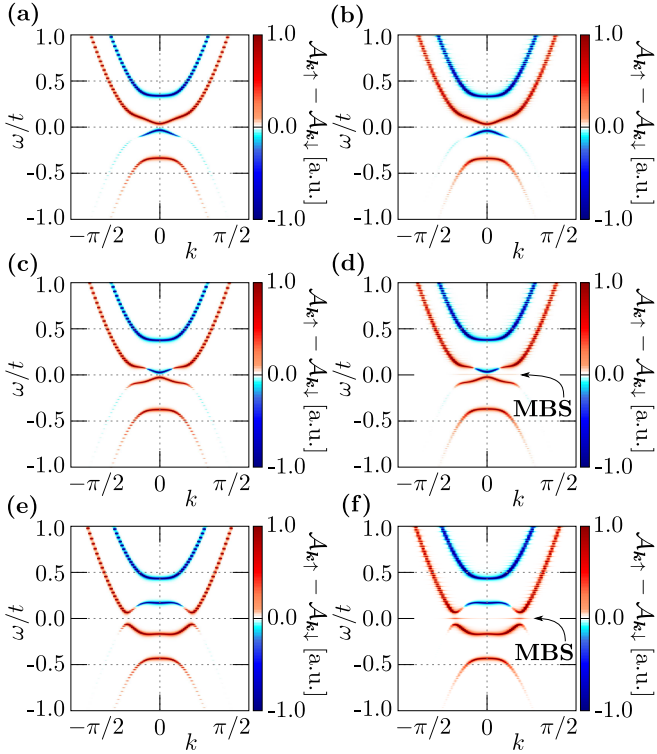}
\caption{
The polarization of the spectral function $\mathcal{A}_{{\bm k}\uparrow} - \mathcal{A}_{{\bm k}\downarrow}$ of the system for different value of the external magnetic field $h / t$ equal $0.15$, $0.2$ and $0.3$ (rows from top to bottoms).
Results for a case of the gate potential $V_{G}$ equal $0 t$ (left column) and $2 t$ (right column).
Topological phase transition occurs for the magnetic field between $0.15 t$ and $0.2 t$.
\label{fig.bands}
}
\end{figure}

Numerical results for spectral function $\mathcal{A}_{{\bm k}\sigma} ( \omega )$ of studied system in the absence of the magnetic field are shown on Fig.~\ref{fig.band_h0}.
In the absence of the QD region [Fig.~\ref{fig.band_h0}(a)], the band structure (blue line) with SOC is shifted and the two bands with clearly seen superconducting gap $\Delta_{0}$ around the Fermi level can be observed.
Additional the QD region (with $\mathcal{N}_{D} = 10$ and $V_{G} / t= 2$) leads to the emergence ABS at the relatively high energy  above the Fermi level ($\omega > V_{G} \pm h$) [Fig.~\ref{fig.band_h0}(b)].
Thus, they can be shown in the total DOS spectrum as a {\it flatbands} of virtual bound states (located outside the gap region)~\cite{balatsky.vekhter.06}.
Additionally, disorder in the system leads to the {\it blurring} of the spectral weight of every state~\cite{medeiros.stafstrom.14,skachkov.quayle.16,sun.xi.17,xu.chiu.17}.
However, the main band structure is still well visible due to the strongest intensitivity of the spectral function.

Now we discuss the influence of polarization of the spectral function $\mathcal{A}_{{\bm k}\uparrow} - \mathcal{A}_{{\bm k}\downarrow}$ in a case of the absence and presence of the QD region (shown in Fig.~\ref{fig.bands} in the left and right column, respectively).
For magnetic field $h < h_{c}$, we observe typical behavior of the topologically trivial phase order of band polarization at the ${\bf k} = 0$ ($\Gamma$ point) -- from upper to lower band with $( \downarrow , \uparrow , \downarrow , \uparrow )$ order [Fig.~\ref{fig.bands}(a) and (b)].
For the critical field $h_{c}$, closing of the gap at the $\Gamma$ point occurs.
Further increase of the magnetic field $h$ leads to the topological phase transition, which is manifested by the band inversion [cf. e.g. Fig.~\ref{fig.bands}(a) and Fig.~\ref{fig.bands}(c)].
We observe a new order of the band polarization within non-trivial topological phase [Fig.~\ref{fig.bands}(c) and (d)] -- $( \downarrow , \downarrow , \uparrow , \uparrow )$.
Further increase of magnetic field $h_{c} \ll h < h_{c2}$  does not change the band polarization sequence [Fig.~\ref{fig.bands}(e) and (f)].
However, we can observe distinct qualitative differences in the band structure [Fig.~\ref{fig.bands}(e) and (f)], while $\Delta_{F} > \Delta_{\Gamma}$.
This characteristic  alterations of the band structure [Fig.~\ref{fig.bands}(c) $\rightarrow$ (e)] are independent of the SOC $\lambda$ and gate potential $V_{G}$.

In the periodic system (without the QD) for $h > h_{c}$, we observe a new gaped nontrivial topological state typical for topological insulator [Fig.~\ref{fig.bands}(c) and (e)].
Due to the existence of ``edge''  of the system, between the ring and the QD region, we can observe a zero--energy dispersionless Majorana flatband [labeled as MBS on panels Fig.~\ref{fig.bands}(d) and (f)].
The MBS emerge as a time reversal invariant electronic state within gapless edge states~\cite{kane.mele.05}.
In this sense, the MBS is realized in the n--dimensional system as a (n-1)--dimensional ``surface'' states~\cite{fu.kane.07,moore.balents.07}.
In our quasi--one--dimensional system, the {\it surface} is given by the zero-dimensional ``edge'' between the QD region and the ring.

\begin{figure}[!b]
\centering
\includegraphics[width=\linewidth]{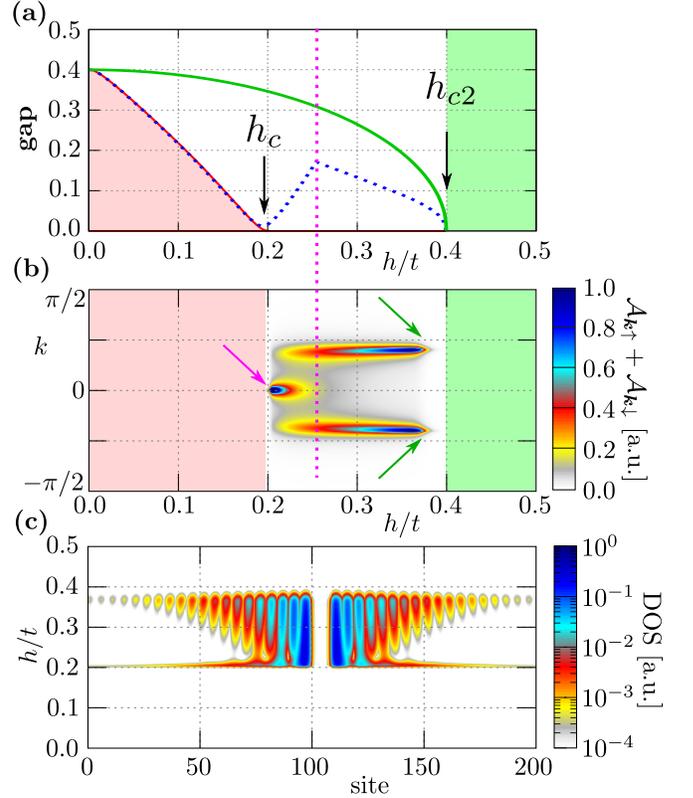}
\caption{
Comparison of the gaps (a), the zero-energy spectral functions $\mathcal{A}_{{\bm k}\uparrow} + \mathcal{A}_{{\bm k}\downarrow}$ (b) and the zero-energy local densities of states LDOS (c) as a function of the external magnetic field $h$.
\label{fig.mbs_mom}
}
\end{figure}

Additionally, the SOC in the system contributes to the helicity basis~\cite{seo.han.12}, in which the non-interacting part of the Hamiltonian $\mathcal{H}$ is diagonal (in the Rashba bands representation).
Then, the superconducting gap has a singlet (interband) and triplet (intraband) pairing components.
Those properties can be observed experimentally~\cite{quay.hughes.10,heedt.traversoziani.17,kammhuber.cassidy.17}.
As a consequence of this, described model can be exactly mapped on to Kitaev model for some range of parameters~\cite{lutchyn.sau.10} (i.e. when $\Delta_{F} \ll \Delta_{\Gamma} $).

\paragraph*{Majorana flatband.}
Now we will discuss the influence of magnetic field $h$ and SOC $\lambda$ on the Majorana flatband (Fig.~\ref{fig.mbs_mom}).

First, we will describe how the gaps change as a function of magnetic field $h$ [Fig.~\ref{fig.mbs_mom}(a)].
For relatively small magnetic field ($0 < h < h_{c}$), the trivial topological phase exist (marked by red region).
An increase of magnetic field leads to the closing of trivial superconducting gap at $h_{c}$~(\ref{eq.qpt_hc}).
For $h = h_{c}$ topological phase transition occurs (see Sec.~\ref{sec.homo}), which is manifested by band inversion (described in previous paragraph).
In the presence of the strong ``defect'' in the form of the QD, system is capable of hosting the topologically protected zero--energy MBS. 
In this state (independent of the boundary condition) the topological gap (shown blue dashed line) strongly depends on the chosen parameter~\cite{sau.tewari.10,sau.lutchyn.10,stanescu.lutchyn.11}.
Further increase of magnetic field $h > h_{c}$ leads to the disappearance of the superconductivity and simultaneous closing of the topological gap at $h_{c2}$.

Using notation described in Fig.~\ref{fig.schem_band}, the topological gap is initially given by $\Delta_{\Gamma}$ and afterwards changed to $\Delta_{F}$~\cite{das.ronen.12} [at the left and right side of the pink doted line at Fig.~\ref{fig.mbs_mom}(a) and (b), respectively].
At the same time, increase of the magnetic $h$ field leads to the modification of the spectral weight in the Majorana flatband [Fig.~\ref{fig.mbs_mom}(b)].
For the case when topological gap is given by $\Delta_{\Gamma}$, we can see significant localization of the spectral weight around ${\bm k} = 0$ (pink arrow).
Similarly, when topological gap is equal to $\Delta_{F}$
the spectral weight is accumulated around ${\bm k} = k_{F}$ (green arrows).
However, we must have in mind that the spectral function is non--zero for any ${\bm k}$.
It is another sign of MBS emergent nature, as the whole band of electrons constitute a single Majorana quasiparticle.

Additionally, change of the spectral function in Majorana band has its reflection in the LDOS [Fig.~\ref{fig.mbs_mom}(c)].
At magnetic field $h_{c}$ we observe maximal delocalization of the zero--energy bound states in space.
Increasing $h$ leads to localization of MBS around the QD boundaries in the form of characteristic LDOS oscillation~\cite{sau.tewari.10,stanescu.lutchyn.11,kobialka.ptok.18}.
When $\Delta_{\Gamma} = \Delta_{F}$ we can observe the maximally localized MBS, what corresponds to the maximal delocalization in momentum space (or in other words, a uniform distribution of the spectral weight).
Described above properties can be explained by the manifestation of Heisenberg's uncertainty principle -- state with clearly defined momentum ${\bm k}$ (in the form of band structure, e.g. Fig.~\ref{fig.schem_band}) at the same time acts in real space as maximally delocalized state.

As a consequence of the properties described above, the MBS cannot be associated with one specific momentum~\cite{lee.lutchyn.17}.
Moreover, in construction of realistic {\it minimal} model of the system with the MBS,one needs to consider both Rashba bands and crossings of the Fermi level around ${\bm k}=0$ and ${\bm k}=\pm k_{F}$~\cite{sticlet.nijholt.17}.

\section{Summary and outlook}
\label{sec.sum}

Localization of the MBS edge state can be probed by the STM~\cite{chevallier.klinovaja.16}.
However, only the spin--resolved measurement of the LDOS~\cite{wiesendanger.09,oka.brovko.14} can be used to distinguish the real MBS from the trivial zero--energy bound state~\cite{sticlet.bena.12,maska.domanski.17}.
This properties of system with the MBS are 	in
agreement with recent expertiments~\cite{jeon.xie.17}.
In similar way, the spin-- and angle-resolved photoelectron spectroscopy (sr--ARPES) can be helpful to investigate the non--topological to topological phase transition measurements~\cite{jozwiak.park.13,zhu.veenstra.14}.
Moreover, described properties of the Majorana flatband in the momentum space can be experimentally measured using the high quality time-- and angle--resolved photoelectron spectroscopy (tr-ARPES).
This type of measurements are sensitive to small details of the unoccupied band structure and recently it has been successfully used in the case of topological insulator WTe$_{2}$~\cite{crepaldi.autes.17} and Bi$_{2}$Se$_{3}$~\cite{soifer.gauthier.17} in a study of the non-trivial topological phase.
In our case of the nanoring with the QD region, the tr-ARPES technique should be helpful to study states around the Fermi level.
Additionally, this type of measurement should allow observation of the evolution of MBS by tracking the intensity node in the circular dichroism~\cite{park.han.12,zhu.veenstra.13,kondo.nakashima.17}.
The joint study using those two techniques (sr-ARPES and tr-ARPES) have been  already performed~\cite{sanchezbarriga.golias.16,bugini.boschini.17}.

Summarizing, we proposed the possibility of the creation and manipulation of the Majorana bound states in the nanoring structure, by the electrostatical formation of the quantum dot region. 
In contrast to the other proposal of realization of the Majorana quasiparticles in the nanoring~\cite{vanmiert.ortix.17},  our method is  non-invasive.
We have shown that the Andreev or Majorana bound states can emerge in the vicinity of electrostatically controlled QD region of nanoring.
The type of induced bound states depends on the value of the potential, and can be easily changed. 
Moreover, from practical point of view, this type of the nanostructure can be prepared relatively simply.

Additionally, we discussed the emergence of the Majorana bound state on the ``induced'' edges in the form of the quantum dot region.
We have shown that the Majorana bound states can be induced in the system together with  Andreev bound states, however, former states are localized in space at the outside of the quantum dot, while the latter -- the ABS, reside inside.
Moreover, we proposed and discussed a method which allows for 
the existence of the Andreev and/or Majorana bound states in the momentum space, and by using it we investigated the properties of the Majorana flatband. 
Our results have shown that the Majorana bound states are associated with the dispersionless zero-energy flatband.
Spectral function analyses shown that those states can not exist within a discrete range of some selected momenta, but should be studied in the context of whole band.
It is a consequence of the leakage of the spectral weight within the Majorana flatband from central momentum point to the Fermi momenta.
Our results should be helpful in a construction of the realistic minimal model of the Majorana bound states.

\begin{acknowledgments}
We thank T. Doma\'n{}ski and Sz. G\l{}odzik for careful reading of the manuscript, valuable comments and discussions.
This work was supported by the National Science Centre (NCN, Poland) under grants  DEC-2014/13/B/ST3/04451 (A.K.) and UMO-2017/26/D/ST3/00054 (A.P.).
\end{acknowledgments}


\bibliography{biblio}

\end{document}